\documentclass[conference]{IEEEtran}
\IEEEoverridecommandlockouts

\usepackage{cite}
\usepackage{amsmath,amssymb,amsfonts}
\usepackage{algorithmic}
\usepackage{graphicx}
\usepackage{textcomp}
\usepackage{xcolor}
\def\BibTeX{{\rm B\kern-.05em{\sc i\kern-.025em b}\kern-.08em
    T\kern-.1667em\lower.7ex\hbox{E}\kern-.125emX}}

\begin{document}

\title{Touch‐Augmented Gaussian Splatting for Enhanced 3D Scene Reconstruction}

\author{%
  Yuchen\,Gao\textsuperscript{a}, 
  Xiao\,Xu\textsuperscript{b}, 
  Eckehard\,Steinbach\textsuperscript{b}, 
  Daniel\,E.\,Lucani\textsuperscript{a}, 
  Qi\,Zhang\textsuperscript{a}\\[-0.2em]
  \textsuperscript{a}\textit{Department of Electrical and Computer Engineering, Aarhus University, Aarhus, Denmark}\\[-0.1em]
  \textsuperscript{b}\textit{TUM School of Computation, Information and Technology, Technical University of Munich, Munich, Germany}\\[-0.1em]
  Email: \{yuchen, daniel.lucani, qz\}@ece.au.dk, \{xiao.xu, eckehard.steinbach\}@tum.de
}

\maketitle

\begin{abstract}

This paper presents a multimodal framework that integrates touch signals (contact points and surface normals) into 3D Gaussian Splatting (3DGS). Our approach enhances scene reconstruction, particularly under challenging conditions like low lighting, limited camera viewpoints, and occlusions. Different from the visual-only method, the proposed approach incorporates spatially selective touch measurements to refine both the geometry and appearance of the 3D Gaussian representation. To guide the touch exploration, we introduce a two-stage sampling scheme that initially probes sparse regions and then concentrates on high-uncertainty boundaries identified from the reconstructed mesh. A geometric loss is proposed to ensure surface smoothness, resulting in improved geometry. Experimental results across diverse scenarios show consistent improvements in geometric accuracy. In the most challenging case with severe occlusion, the Chamfer Distance is reduced by over 15$\mathbf{\times}$, demonstrating the effectiveness of integrating touch cues into 3D Gaussian Splatting. Furthermore, our approach maintains a fully online pipeline, underscoring its feasibility in visually degraded environments. 

\end{abstract}


\section{Introduction}
\label{section:introduction}

Across diverse fields, from robotic manipulation \cite{b1} and industrial inspection \cite{b2} to the emerging Tactile Internet (TI) \cite{b3,b4}, there is a growing need to transcend the physical boundaries by building a 3D model of the environment. It allows users to perceive and interact with hazardous or inaccessible scenes. The core of this need lies in fast and reliable 3D reconstruction that strikes a balance among geometric accuracy, robustness, and time. Unlike classical point-to-point teleoperation \cite{b5,b6}, these applications prioritize holistic environmental understanding, both visually and geometrically. While vision-based methods form the backbone of such tasks, visual-only methods are unreliable when image data is incomplete. To be specific, holistic environmental understanding must address three key challenges: (1) highly efficient incremental updates; (2) robustness to lighting/occlusion degradations; (3) adaptive refinement of uncertain regions.

To meet these challenges, the fidelity and efficiency of a 3D reconstruction method play critical roles. However, traditional point cloud-based approaches \cite{b11} lack surface topology, especially when the input is sparse, and the visual quality can be compromised. While mesh-based approaches accurately capture the geometry and physics of the model, they are not flexible
in updates, and remeshing can be costly \cite{b19}. Meanwhile, advanced neural implicit representations such as NeRF \cite{b12} and InstantNGP \cite{b13} enable photorealistic rendering but may struggle to adapt quickly to dynamic changes.  

In contrast, 3D Gaussian Splatting (3DGS) \cite{b10} employs anisotropic Gaussian primitives to represent scenes explicitly while supporting differentiable optimization. We choose 3DGS because (i) each Gaussian can be inserted or pruned in $O(1)$, matching per-touch online update loop in Section~\ref{section:framework}, (ii) the analytical covariance gives a closed-form directional radius in Eq.~\eqref{eq:radius} that our geometric loss differentiates cheaply to guide incremental updates. However, like all vision-centric methods, 3DGS depends on photometric consistency. Sparse yet reliable touch points resolve these ambiguities. While visual information is easy to obtain with today's wealth of camera solutions, it may miss fine details. In contrast, touch information is more accurate but less efficient to acquire.

While recent vision–touch studies show that multimodal cues help basic manipulation and offline shape completion, they seldom model environmental degradation due to poor lighting, occlusion, or limited viewpoints. Early teleoperation-oriented approaches by Xu \emph{et al.}~\cite{b7,b8} ensured haptic stability through a local proxy model, but paid less attention to fine-grained scene geometry.  
Antonsen \emph{et al.}~\cite{b9} fused contact data into a global signed-distance field, incurring expensive full-volume updates.  
More recent systems exploit high-resolution tactile skins or task-specific pipelines: TouchFusion~\cite{b14} fuses GelSight measurements with RGB-D offline. Yi \emph{et al.}~\cite{b15} and Ottenhaus \emph{et al.}~\cite{b16} drive an active probe with Gaussian-process or implicit-surface uncertainty, while Rustler \emph{et al.}~\cite{b17} focus on shape completion for grasp success.  
All rely on special hardware or offline pipelines, so they are not online-ready.

Crucially, the above work is tuned for task completion in teleoperation (i.e., maximizing grasp or insertion success), whereas our goal is to deliver a visually, metrically accurate environment model that users can inspect even when vision degrades. Most touch-aided systems (e.g. TouchFusion~\cite{b14}, Yi \emph{et al.}~\cite{b15}) optimize these task-level rewards and run offline. None of those systems provides guarantees on reconstruction fidelity under poor lighting, occlusion, or restricted viewpoints, a gap that our touch-augmented 3DGS framework tries to bridge. Thus, direct metric-wise comparison would be misleading.

\begin{figure*}[htbp]
    \centering
    \includegraphics[width=.6\textwidth]{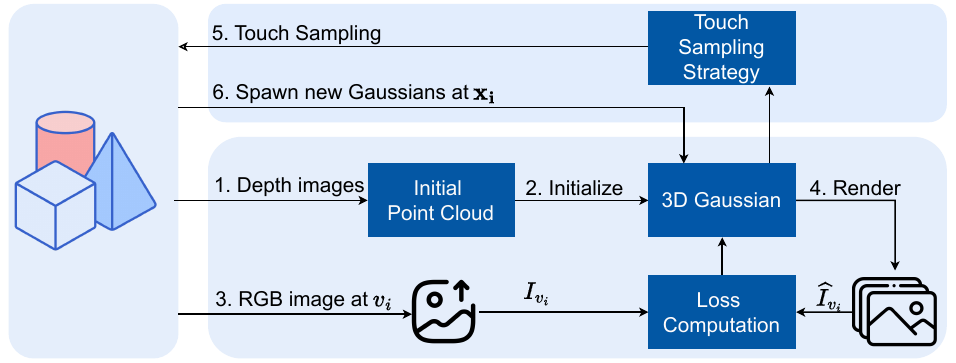}  
    \caption{Overview of the touch-augmented 3D Gaussian Splatting framework. (1) Depth cameras capture images to generate an initial point cloud (2). This point cloud is then represented using Gaussian ellipsoids. RGB cameras acquire training images \(I_{v_i}\) (3), compared against the rendered images \(\hat{I}_{v_i}\) (4). A touch sampling strategy (5) refines the Gaussian model in areas of sparse coverage or high uncertainty, spawning new Gaussians at \(\mathbf{x_i}\) where contact points and normals are acquired (6). The overall loss function consists of both an image-based loss (comparing \(I_{v_i}\) and \(\hat{I}_{v_i}\)) and a geometric loss (enforcing local surface consistency from the contact points and normals), which improve the 3D reconstruction.}

    \label{fig:pipeline}
\end{figure*} 

The main contributions are summarized as follows:
\begin{itemize}
    \item Introduction of an online training pipeline that integrates 3DGS with selective touch patches to incrementally refine geometry, enhancing the reliability of visual-based reconstruction under hazardous environments.
    \item Design of an iterative touch sampling strategy that identifies under-reconstructed regions and injects localized touch measurements, enforcing consistency among neighboring Gaussian primitives.
      
\end{itemize}
The proposed method achieves a maximum of a 15$\times$ reduction in the Chamfer Distance (CD), improves the F-score by as much as 43\%, and decreases the Jensen-Shannon Divergence (JSD) by up to 88\% across various scenarios. It demonstrates the feasibility of rendering a comprehensible scene in an environment with poor visibility.
\begin{table*}[ht]
    \centering
    \caption{Notation Table}
    \label{tab:notation}
    \renewcommand{\arraystretch}{1.2}
    \small  
    \begin{tabular}{p{0.9cm} p{3cm} p{0.9cm} p{3cm} p{0.9cm} p{3cm} p{0.9cm} p{3cm}}
        \hline
        \textbf{Notation} & \textbf{Meaning} & \textbf{Notation} & \textbf{Meaning} & \textbf{Notation} & \textbf{Meaning} & \textbf{Notation} & \textbf{Meaning} \\
        \hline
        $\mathbf{\mu}$ & Mean of a Gaussian & $\mathbf{x}$ &points on Gaussian& $\mathbf{\Sigma}$ & Covariance matrix & $\mathcal{G}(\mathbf{x})$ & Gaussian function\\ 
        $\mathbf{R}$ & Rotation matrix& $\mathbf{S}$ & Scaling matrix & $\mathbf{\Sigma}_p$ & Projected covariance & $\mathbf{J}$ & Jacobian \\ 
        $\mathbf{W}$ & Viewing transformation &$\alpha_i$ & Opacity of Gaussian $i$ & $\mathbf{c}_i$ & Color contribution & $T_i$ & Transmittance \\
        $\sigma_i$ & Density of Gaussian $i$ & $\delta_i$ & Distance interval & $\mathbf{c}$ & Final rendered color & $\mathbf{R}_c^w, \mathbf{t}_c^w$ & Rotation \& Translation \\ $\mathbf{T}_c^w$ & Camera-to-world & $\mathbf{p}$ & Point in world axes & $\mathbf{p}_c$ & Point in camera axes & $(u, v)$ & Pixel coordinates \\
        $z$ & Depth & $f_x, f_y$ & Focal lengths & $c_x, c_y$ & Principal point
        & $\mathcal{L}_{\text{image}}$ & Visual loss\\ 
        $\mathcal{L}_{\text{touch}}$ & Touch loss &$\lambda$ & Loss weight & $N$& Number of pixels & $\mathbf{v}_{ij}$ &Directional Vector  \\ 
        $a,b,c$ & Ellipsoid principal axes &$d_{ij}$ & Euclidean distance& $r_i, r_j$ & Directional radius&$\mathbf{M}_i$ & Ellipsoid matrix\\
        $\mathcal{P}_i$ & Touch patch& $\mathcal{Q}_i$ & $k$-nearest point set &$v_i$ & $i$-th view angle &$\hat{I}_i$ & Rendered image \\
        $ I_i$ & Ground truth image &$\mathcal{R}$ & Rendering function &\( \|\cdot\|\)& Euclidean distance  \\
        \hline
    \end{tabular}
\end{table*}


\section{Background}
\label{section:background}
We will explain the fundamentals of 3DGS and touch sampling in this section.
 
\subsection{3D Gaussian Splatting}

\subsubsection{Mathematical Representation}
3DGS is a differentiable rendering technique that represents three-dimensional scenes using Gaussian primitives. Each Gaussian is defined by its center \(\mathbf{\mu} \in \mathbb{R}^3\) and its covariance matrix \(\mathbf{\Sigma} \in \mathbb{R}^{3 \times 3}\), which encodes the spatial extent and anisotropic shape of the Gaussian. The Gaussian function is expressed as \cite{b8}:
\begin{equation}
\mathcal{G}(\mathbf{x}) = \exp\left( -\frac{1}{2} (\mathbf{x} - \mathbf{\mu})^\top \mathbf{\Sigma}^{-1} (\mathbf{x} - \mathbf{\mu}) \right) ,
\end{equation}
where \(\mathbf{x}\) is the coordinate of the points on Gaussian's surface.

\subsubsection{Covariance Matrix Parameterization}

To ensure that the covariance matrix \(\mathbf{\Sigma}\) remains positive semi-definite during optimization, it is parameterized as \cite{b8}: \(\mathbf{\Sigma} = \mathbf{R} \mathbf{S} \mathbf{S}^\top \mathbf{R}^\top,
\) where: \(\mathbf{S} \in \mathbb{R}^{3 \times 3}\) is a diagonal scaling matrix that determines the anisotropic spread of the Gaussian. \(\mathbf{R} \) is a Special Orthogonal (\(SO\)(3)) rotation matrix representing the orientation.


\subsubsection{Projection and Rendering}

To render 3D Gaussians in 2D, the covariance matrix \(\mathbf{\Sigma}\) is transformed into the camera coordinate system using a viewing transformation \(\mathbf{W}\). The resulting covariance matrix \(\mathbf{\Sigma_p}\) is:
\begin{equation}
\mathbf{\Sigma}_p = \mathbf{J} \mathbf{W} \mathbf{\Sigma} \mathbf{W}^\top \mathbf{J}^\top ,
\end{equation}
where \(J\) is the Jacobian of the projective transformation. 

The rendering corresponds to step 4 in Fig.~\ref{fig:pipeline}, it employs \(\alpha\)-blending, where the color \(\mathbf{c}\) of a pixel is computed as \cite{b8}:
\begin{equation}
\mathbf{c}= \sum_{i=1}^{N} T_i \alpha_i \mathbf{c_i} ,
\end{equation}
where \(\alpha_i = 1 - \exp(-\sigma_i \delta_i)\) is per-sample opacity (derived from density $\sigma_i$), and the transmittance is \(T_i = \prod_{j=1}^{i-1}(1 - \alpha_j)\). Color \(\mathbf{c}\) is taken along the ray with intervals \(\delta\). \(\mathbf{c_i}\) is the color contribution of the \(i\)-th Gaussian. 

\subsection{Touch sampling strategy}
Haptic feedback has been widely explored in digital reconstruction and modeling to enhance geometry understanding. The key is to decide where to touch in terms of both locating the object and deciding where the next sample patch should be~\cite{b14,b15,b16,b17}.
Previous works proposed mainly two ways of sampling decisions: uncertainty-driven solutions and random selection~\cite{b14}. Despite the fact that random selection is easier, it is less efficient and less explainable than the uncertainty-driven solutions.
For example, Z. Yi \emph{et al.} \cite{b15} model the surface using the Gaussian process to decide the uncertain region to sample. 
S. Ottenhaus \emph{et al.}~\cite{b16} utilize a constantly updated implicit surface, and sampling points are dynamically adjusted to prioritize the exploration of uncovered or error-prone regions. L. Rustler \emph{et al.}~\cite{b17} extend the concept of implicit surface guiding sampling to incorporate both touch and visual for more efficient information extraction and a higher success rate of task completion. 

\section{TOUCH-AUGMENTED 3D GAUSSIAN SPLATTING}
\label{section:framework}
In this section, we detail the workflow of touch-augmented 3DGS framework. Starting from camera and sensor setup, we will describe online training procedure, including touch patch integration and boundary-based sampling. The description is shown in Fig.~\ref{fig:pipeline}. The framework starts from an initialized point cloud. By expanding each point into an ellipsoid, each point contains more information (e.g., opacity, rotation). The Gaussian primitives are rendered through rasterization. 
\subsection{Data acquisition}
Required raw data includes depth images, RGB images and touch patches. The visual data is acquired during steps 1-3 in Fig.~\ref{fig:pipeline}, and touch samples are retrieved during step 6.
\paragraph{Visual data acquisition}
Point clouds are generated from depth images by projecting the depth measurements using camera intrinsics in step 1 in Fig.~\ref{fig:pipeline}. Each point \(x_c, y_c, z_c\) is derived from the pixel column \((u, v, z)^T\). Depth images are transformed into 3D point clouds using intrinsic parameters \( (f_x, f_y, c_x, c_y) \) by considering that:
\begin{equation}
\begin{aligned}
x_c &= \frac{(u - c_x) \cdot z}{f_x},\quad y_c &= \frac{(v - c_y) \cdot z}{f_y},\quad z_c &= z 
\end{aligned}
\end{equation}
 where \(f_x\) and \(f_y\) are the focal lengths, and \(c_x\) and \(c_y\) represent the camera center coordinates. The point in camera space \(\mathbf{p}_c\) is then transformed to the world coordinate system using the known rigid body transformation matrix,\(\mathbf{T}_c^w= \begin{bmatrix}
\mathbf{R}_c^w & \mathbf{t}_c^w \\
\mathbf{0} & 1
\end{bmatrix}  
\), and the pixel in world space is given by \begin{equation}
 \begin{bmatrix}
     \mathbf{p}\\1
 \end{bmatrix} = \mathbf{T}_c^w  
\begin{bmatrix}
\mathbf{p}_c\\ 1   
\end{bmatrix} ,
\end{equation}
where \(\mathbf{p}\) denotes the point in the world space. Similarly, the image data is transformed in step 2 in Fig.~\ref{fig:pipeline} through standard coordinate translation procedures as the point cloud.  
\paragraph{Touch acquisition}
During touch sample acquisition, we sample the Unified Robot Description Format (URDF) mesh directly: at each iteration, a virtual probe emulating a parallel-jaw (two-finger) gripper selects the two most sparse Gaussian centers. Let $k$ be the number of nearest neighbors (here, $k=400$). We then build a k-dimensional (k-d) tree to retrieve the $k$ nearest mesh points to each center, forming a touch patch \(\mathcal{P}_i = \{(\mathbf{p}_{ij}, \mathbf{n}_{ij}) \mid \mathbf{p}_{ij} \in \mathcal{Q}_i\}\). This simulates dual end-effector contacts on the model, providing local position and normal information for geometric optimization.

This structure facilitates rapid querying of neighboring points within specified spatial constraints. For each sampling center \( k \)-nearest neighbors from the ground truth point cloud are retrieved, forming a touch patch \(\mathcal{P}_i = \{ (\mathbf{p}_{ij}, \mathbf{n}_{ij}) \mid \mathbf{p}_{ij} \in \mathcal{Q}_i \}\), where \( \mathcal{Q}_i = \{\mathbf{p}_{i1}, \mathbf{p}_{i2}, \dots, \mathbf{p}_{ik}\} \) represents the set of \( k \)-nearest ground truth points to each sampling center.

\subsection{Training}
Data flow consists of two modalities, visual and touch, which are interleaved and fed into the training of the Gaussian model.
\subsubsection{Visual}
The training pipeline initializes the scene and Gaussian model of the object (Fig.~\ref{fig:pipeline} step 2) based on the point cloud from the depth cameras (Fig.~\ref{fig:pipeline} step 1). Images are taken to further train the Gaussian model (Fig.~\ref{fig:pipeline} step 3). The optimization of the parameters of the Gaussians is based on the forward and backpropagation. Gaussians \(\mathcal{G}\) are rendered in parallel on the GPU through rasterization at the corresponding view angle. The Gaussians are optimized by calculating the difference between the rendered image \( \hat{I}_i \) and ground truth \( I_i \). The discrepancy between the rendered image \( \hat{I}_i \) and the ground truth image \( I_i \) is quantified using a composite visual loss function~\cite{b8}:

\begin{equation}
    \mathcal{L}_{\text{image}} = (1 - \lambda ) \frac{1}{N} \sum_i \| \hat{I}_i - I_i \|_1 
    + \lambda \left( 1 \hspace{-.5mm}-\hspace{-.5mm} \text{SSIM}(\hat{I}_i, I_i) \right).
\end{equation} 

The \( \| \hat{I}_i - I_i \| \) measure is the absolute difference between the rendered and ground truth images. \( \text{SSIM}(\hat{I}_i, I_i) \) denotes the Structural Similarity Index Measure (SSIM), assessing the perceived quality and structural similarity between the two images. \( \lambda \) is a weighting factor that balances the contributions. Despite these advantages, 3DGS still suffers from low light and occlusion, motivating the touch cue we introduce next.

\subsubsection{Touch}
 \label{touch-sampling}
We have two different strategies (Fig.~\ref{fig:pipeline} step 5), Sparsity stage and Boundary stage:
\begin{itemize}
  \item \textbf{Sparsity stage:} Identify the two Gaussians with the largest nearest-neighbor gaps. For each, sample their \(k\) nearest mesh points and create new Gaussians there whose centers will no longer be adjusted.
  \item \textbf{Boundary stage:} We intermittently build a temporary Poisson-surface proxy solely to reveal boundary holes, leveraging the mesh’s geometric accuracy while retaining the Gaussians’ efficiency and flexibility for subsequent refinement.
\end{itemize}

\paragraph{Sparsity-based sampling}
Since robotic manipulation tasks often involve bimanual interactions or multiple contact points (e.g., two-fingered grasping). We therefore take two contact regions as an example in our experiments. Although our probe is virtual, limiting each iteration to two patches faithfully emulates common parallel-jaw or two-fingered grasps and keeps the added computation bounded. Our sparsity-based sampling method identifies the two most sparse regions in the current Gaussian model by computing nearest neighbor distances. For each Gaussian \( G_i \), calculate the distance to its nearest neighbor \( d_i \). Gaussians are sorted in descending order of sparsity, and the top \( k \) are selected as sampling centers. 
 
\paragraph{Uncertainty-based sampling} After sparsity-based sampling, we build a temporary Poisson proxy solely to expose boundary holes. It is discarded right after identifying uncertain regions. The mesh edges pinpoint high-uncertainty holes or unreliable parts, steering the next touch patch to geometry that vision alone undersamples. Because this fit is invoked only intermittently, it adds negligible overhead while providing a more reliable uncertainty cue than point-wise density heuristics.

Mesh edges are identified by analyzing the connectivity and normal consistency of the mesh vertices. An edge represents the regions with discontinuity. These boundary points are organized using a k-d tree to accelerate inquiry. A greedy search will be applied to select the coverage centers that maximize the coverage of uncovered boundary points. By iteratively choosing the center, the algorithm ensures coverage of uncertainty regions. From the newly sampled touch points, new Gaussians are spawned based on the touch information and are locked. Conversely, the visually spawned Gaussians nearby are recognized as inaccurate Gaussians. They are pruned to maintain model fidelity.

At each optimization step, we deliberately sample two new surface patches, one per fingertip of a virtual parallel-jaw gripper to match the budget of a realistic two-finger grasp. Across successive iterations these patch pairs are reselected in the most uncertain regions, so the set of contacted areas gradually expands over the whole object.

\paragraph{Geometry loss}
The geometric loss is defined using two basic principles: minimizing the distance between two neighboring Gaussians and minimizing their overlap. The touch loss \(\mathcal{L}_{\text{touch}} \) quantifies the discrepancy between the actual distances between the centroids of two Gaussians and the expected distances based on their directional radius. Defining the vector \( \mathbf{v}_{ij} = \mathbf{\mu}_i - \mathbf{\mu}_j \) that points from Gaussians \( \mathcal{G}_i \) to \( \mathcal{G}_j \), for each pair \( (\mathcal{G}_i, \mathcal{G}_j) \), the directional radius \( r_i \) is the distance from the centroid to the surface of $\mathcal{G}_i$ along $\mathbf{v}_{ij}$.

Let Euclidean distance \( d_{ij} = \|\mathbf{v}_{ij} \| \). The touch loss for each pair \( (\mathcal{G}_i, \mathcal{G}_j) \), $\delta_{ij}$, is defined:
\begin{equation}
    \delta_{ij} = d_{ij} - (r_i+ r_j).
\end{equation}
 
Consider an axis-aligned ellipsoid centered at the origin with semi-axes lengths \(a, b, c\). Its implicit equation can be expressed using the inverse covariance matrix:
\begin{equation}
\label{eq:canonical_form}
    \mathbf{x}^T \mathbf{\Sigma}^{-1} \mathbf{x} = 1 ,
\end{equation}
where $\Sigma^{-1} = \mathrm{diag}(a^{-2},\,b^{-2},\,c^{-2})$ is a diagonal matrix, encodes the ellipsoid's shape.

 For a rotation matrix \(\mathbf{R} \in SO\)(3) (satisfying \(\mathbf{R}^{-1} = \mathbf{R}^T\)), the transformed coordinates \(\mathbf{x}' = \mathbf{R}\mathbf{x}\) yield:
\begin{equation}
    (\mathbf{R}^T\mathbf{x}')^T \mathbf{\Sigma}^{-1} (\mathbf{R}^T\mathbf{x}') = 1 \quad \hspace{-3mm}\Rightarrow \hspace{-3mm} \quad \mathbf{x}'^T (\mathbf{R}\mathbf{\Sigma}^{-1}\mathbf{R}^T) \mathbf{x}' = 1 .
\end{equation}
For a Gaussian \(\mathcal{G}_i\) centered at \(\mathbf{\mu}_i\) with rotation \(\mathbf{R}_i\), its world-space representation becomes:
\begin{equation}
    (\mathbf{x} - \mathbf{\mu}_i)^T \underbrace{\mathbf{R}_i\mathbf{\Sigma}_i^{-1}\mathbf{R}_i^T}_{\mathbf{M}_i} (\mathbf{x} - \mathbf{\mu}_i) = 1 ,
\end{equation}
where \(\mathbf{M}_i\) is precomputed to accelerate. 

The directional radius can be calculated directly using the transformed covariance:
\begin{equation}
\label{eq:radius}
    r_i = \frac{1}{\sqrt{\mathbf{v}_{ij}^T \mathbf{M}_i \mathbf{v}_{ij}}} = \frac{1}{\sqrt{\mathbf{v}_{ij}^T (\mathbf{R}_i\mathbf{\Sigma}_i^{-1}\mathbf{R}_i^T) \mathbf{v}_{ij}}} .
\end{equation}

\section{PERFORMANCE EVALUATION}
\label{section:experiment}
\begin{figure*}[h]
    \centering
    \includegraphics[width=0.78\linewidth]{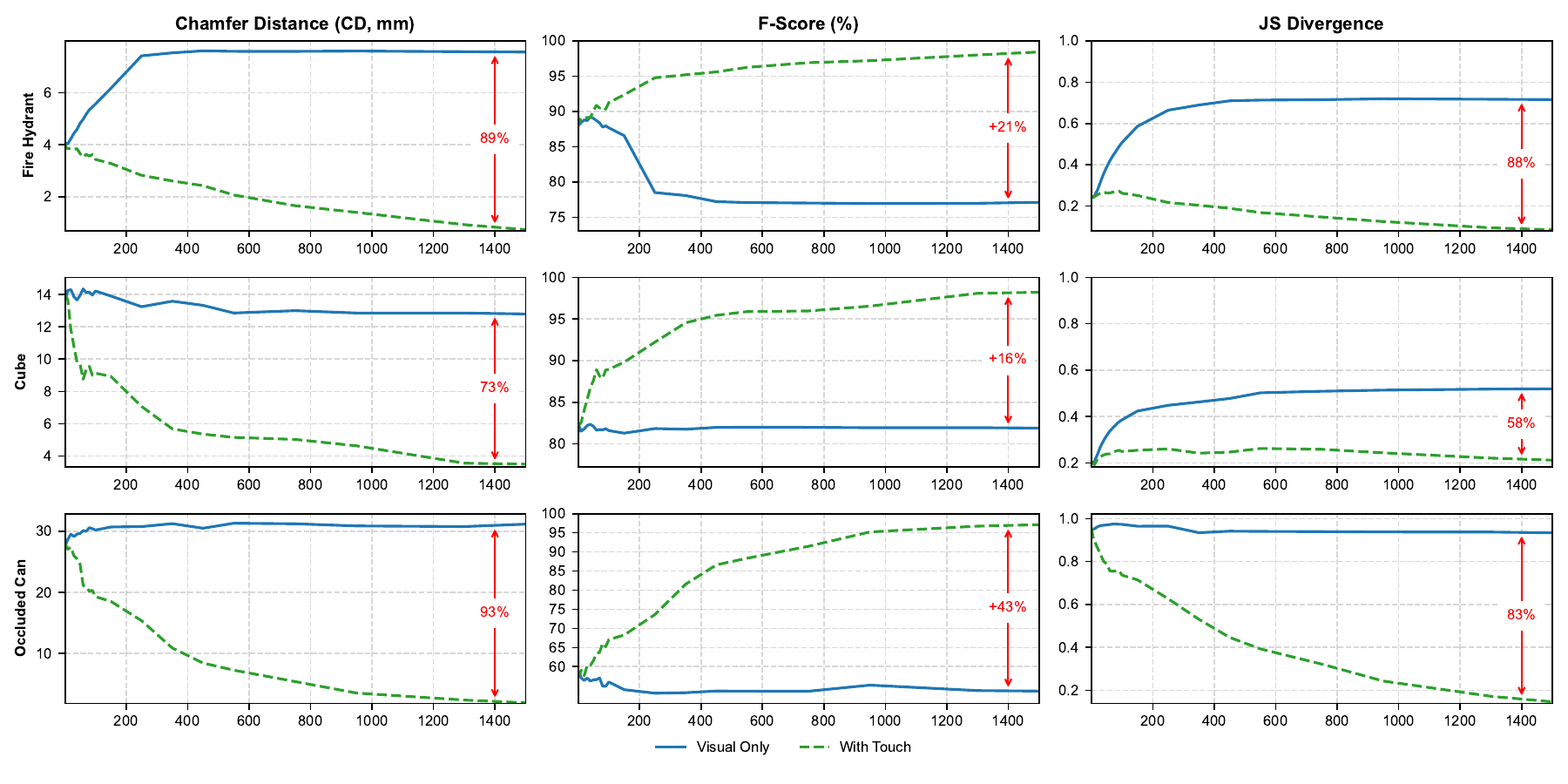}\\[0.5em]
   \caption{%
        Reconstruction quality over training iterations in \textbf{three \emph{separate}} challenging scenarios.
        \textbf{Rows}: (1) \emph{Fire-Hydrant} under \textit{deteriorated lighting}; (2) \emph{Cube} with \textit{missing camera viewpoints}; (3) \emph{Can} under \textit{severe occlusion}.
        \textbf{Columns}: Chamfer Distance (CD, mm), F-score (\%), and Jensen–Shannon Divergence (JSD).
        (Solid blue is the visual-only baseline; dashed green is the proposed touch-augmented method;
        Red arrows is the relative improvement.)%
    }
    \label{fig:comparisons}
\end{figure*}
\subsection{Experimental setup}
The experimental setup was designed to evaluate the effectiveness of touch-aided Gaussian splatting for improving 3D reconstruction in environments with poor visibility. A set of virtual cameras (both depth and RGB) was positioned around the target object to ensure spatial coverage. In particular, four depth cameras were placed at different viewpoints around the object. Nine RGB cameras were placed in overhead and corner positions, providing supplementary visual information. To simulate harsh visual scenarios, the objects were placed in a controlled environment that allowed for variable lighting conditions and introducing occlusions. The experiments include three objects (i.e., Fire Hydrant, Cube, and Can) under the three harsh visual conditions: (1) under deteriorated lighting (i.e., using optimal lighting for point cloud initialization with nine light sources evenly distributed above the object and camera, and employing intentionally poor lighting for 3D reconstruction training with only one light source positioned above the object and no side illumination); (2) with the coverage points missing because some camera views are unavailable; (3) occlusion with three blocks.
We deliberately pick three mutually-distinct shapes to span the typical curvature spectrum. Because our algorithm is shape-agnostic, running it on additional models yielded the same relative improvements. We therefore report these representative results.

\begin{figure}
\centerline{\includegraphics[width=0.33\textwidth]{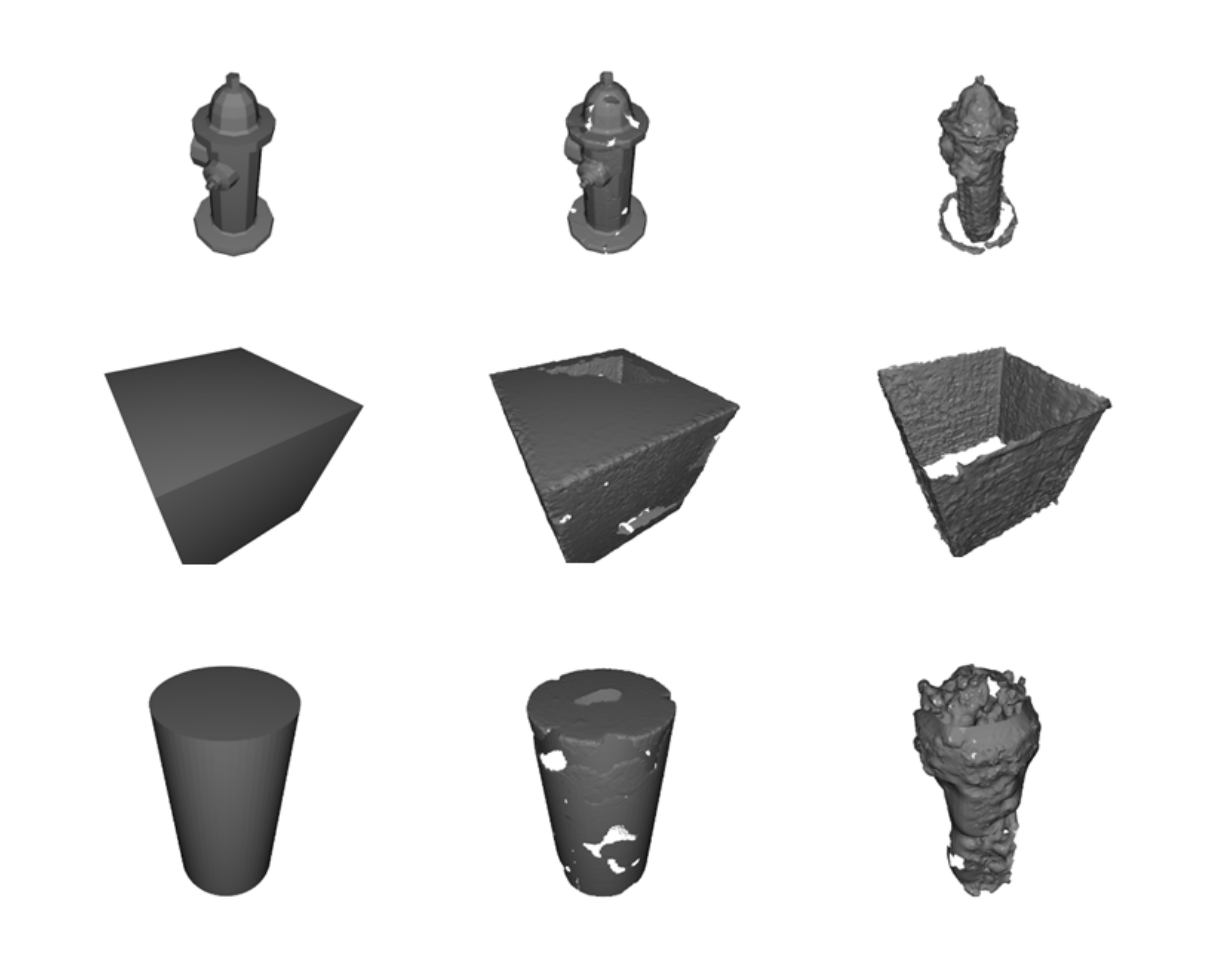}}
\caption{%
    Qualitative reconstructions at iteration 1400.  
    For each object (top to bottom: Fire-Hydrant, Cube, Can) we show, from left to right:
    \textbf{(i)} ground-truth mesh,  
    \textbf{(ii)} touch-augmented result, and  
    \textbf{(iii)} visual-only baseline.  
    Small gaps visible in the second column arise because our method actively prunes Gaussians whose geometry is contradicted by the latest contacts. Those regions are re-filled once they are revisited by later touch samples. 
    }
\label{fig:meshimages}
\end{figure}
We conducted experiments using Gazebo \cite{b18}. Object models were imported, each described in COLLADA format. To obtain high-fidelity ground-truth data, we sampled the COLLADA files and exported dense point clouds that represent the object's geometry. This ground-truth sampling serves as a reference to measure the accuracy of the reconstructed models. To facilitate data acquisition and system integration, each virtual camera in Gazebo was connected to a Robot Operating System (ROS) framework. Specifically, each camera publishes images (or depth images) that are stored as ground truth for training the 3DGS model. 

\subsection{Experimental Results and Evaluation}
We evaluated the reconstruction quality using three complementary metrics. \textbf{Chamfer Distance} (CD) quantifies the geometric discrepancy between the reconstructed point cloud and the ground truth. Lower CD values indicate a more accurate reconstruction. \textbf{F-score} provides a balance between precision and recall for 3D reconstructions. Higher F-scores suggest a more complete and accurate surface recovery. \textbf{Jensen-Shannon Divergence} (JSD) measures the dissimilarity between the reconstructed and ground-truth shape distributions. In this paper, JSD is computed with a base-2 logarithm, ranging from $[0,1]$. Lower JSD values indicate that the reconstructed geometry is better aligned with the real object. 
\begin{table*}[htbp]
    \caption{Performance Metrics at Iteration 750 (CD in mm, F-score in \%, JSD)}
    \centering
    \renewcommand{\arraystretch}{1.2} 
    \small  
    \begin{tabular}{|c|ccc|ccc|ccc|}
        \hline
        \textbf{Case} & 
        \multicolumn{3}{c|}{\textbf{Deteriorated Light}} & 
        \multicolumn{3}{c|}{\textbf{Missing View Angle}} & 
        \multicolumn{3}{c|}{\textbf{Occlusion}} \\
        \cline{2-10}
         & \textbf{CD} & \textbf{F-score} & \textbf{JSD} & 
         \textbf{CD} & \textbf{F-score} & \textbf{JSD} & 
         \textbf{CD} & \textbf{F-score} & \textbf{JSD} \\
        \hline
        \textbf{Fire Hydrant (Visual Only)} & 
            7.6 mm & 77.02\% & 0.7151 & 
            6.7 mm & 83.59\% & 0.7135 & 
            5.7 mm & 89.23\% & 0.5557 \\
        \hline
        \textbf{Fire Hydrant (With Touch)} & 
            1.68 mm & 96.92\% & 0.1461 & 
            1.82 mm & 96.10\% & 0.1520 & 
            1.68 mm & 97.33\% & 0.1452 \\
        \hline
        \hline
        \textbf{Cube (Visual Only)} & 
            10.2 mm & 92.93\% & 0.5955 & 
            13.0 mm & 81.98\% & 0.5094 & 
            18.0 mm & 73.21\% & 0.6021 \\
        \hline
        \textbf{Cube (With Touch)} & 
            4.71 mm & 95.82\% & 0.2210 & 
            5.02 mm & 95.96\% & 0.2595 & 
            6.24 mm & 88.97\% & 0.2575 \\
        \hline
        \hline
        \textbf{Can (Visual Only)} & 
            12.1 mm & 25.65\% & 0.9173 & 
            13.8 mm & 20.64\% & 0.9630 & 
            31.21 mm & 53.57\% & 0.9393 \\
        \hline
        \textbf{Can (With Touch)} & 
            4.35 mm & 78.36\% & 0.4496 & 
            4.73 mm & 72.58\% & 0.4826 & 
            5.43 mm & 91.46\% & 0.3220 \\
        \hline
    \end{tabular}
    \label{tab:perfomat750}
\end{table*}

Table~\ref{tab:perfomat750} compares the reconstruction performance results of the visual-only and touch-augmented methods for three objects (Fire Hydrant, Cube and Can) under three test conditions measured at iteration 750 of training. The results show that integrating contact points and normals improve geometric accuracy across all test scenarios. The extra computation stems mainly from (i) twice KD-tree range queries at \(O(k\log N)\) (k = 400, N $\leq 5\times10^5$) and (ii) the optimization of the newly spawned Gaussians ($\leq$ 1 \%). Empirically, this method introduces less than 20\% overhead compared to the visual-only runtime on a modern GPU, so no noticeable slowdown was observed during training.

Fig.~\ref{fig:comparisons} shows the reconstruction performance improvement with more iterations. This figure shows the experimental results of (1) Fire Hydrant under deteriorated lighting (i.e., with good light condition for initializing point cloud but poor light condition during training for 3D reconstruction); (2) Cube with the coverage points missing due to missing some of the camera views; (3) Can with heavy occlusion.

Quantitative results in Fig.~\ref{fig:comparisons} highlight these advancements through three key metrics: Chamfer Distance (CD), F-score, and Jensen-Shannon Divergence (JSD). Taking an example of iteration at 1400, for the Fire Hydrant case under deteriorated lighting, the touch-augmented method improves the CD by 89\% (from 7~mm to 0.7~mm), the F-score by 21\% (from 76\% to 97\%), and JSD by 88\% (from 0.7 to below 0.1). Similarly, in the experiment of the Cube with missing views at iteration 1400, the CD improves by 73\% (from 13~mm to 3.5~mm), accompanied by a 16\% increase in F-score (from 81\% to 97\%), and JSD by 58\% (from 0.5 to 0.21). The improvement of reconstruction performance is most significant in the scenario under severe occlusion. From Fig.~\ref{fig:comparisons}, it shows that at iteration 1400, the visual-only method reaches a CD around 30~mm and an F-score below 60\%, integrating touch modality drastically reduces the CD from 30~mm to 2~mm (15$\times$) and boosts the F-score above 95\%, and JSD by 83\% (from 0.9 to lower than 0.2).

These improvements can also be reflected by qualitative analysis. As illustrated in Fig.~\ref{fig:meshimages}, the touch-augmented method shows enhanced geometric details in visually ambiguous regions. For instance, the bolts on the fire hydrant and ridges of the can are lost in vision-only reconstructions; In contrast, they are accurately captured despite adverse conditions. Touch information directly compensates for missing view information or occluded areas, enabling the framework to resolve ambiguities where visual data is sparse or unreliable. Both quantitative and qualitative analysis illustrate the feasibility of the touch-augmented approach in refining 3D reconstructions in scenarios where traditional vision-based methods fail.

\section{Conclusion}
\label{section:conclusion}
We introduced an online touch-augmented 3DGS that preserves geometry when vision fails. By integrating touch modality into the 3DGS training process, our framework effectively completes visual data, ensuring more robust geometric modeling even under challenging conditions such as poor lighting, limited camera viewpoints, and severe occlusions. The proposed method employs a two-stage touch sampling strategy, ensuring targeted and efficient refinement of the 3D model. Our study highlights the potential of multimodal data fusion in advancing 3D geometric modeling, particularly in environments where traditional vision-based techniques are insufficient. This touch-augmented 3DGS framework offers a robust solution for 3D perception in robotics, teleoperation, and other settings demanding high-fidelity geometry under adverse visual conditions. 
\section*{Acknowledgments}
This research was supported by the TOAST project, funded by the European Union’s Horizon Europe research and innovation program under the Marie Skłodowska-Curie Actions Doctoral Network (Grant Agreement No. 101073465), the Danish Council for Independent Research project eTouch (Grant No. 1127-00339B), and NordForsk Nordic University Cooperation on Edge Intelligence (Grant No. 168043).

\end{document}